\documentclass{article}

\usepackage{subcaption}
\usepackage{arxiv}

\usepackage[utf8]{inputenc} 
\usepackage[T1]{fontenc}    
\usepackage{hyperref}       
\usepackage{url}            
\usepackage{booktabs}       
\usepackage{amsfonts}       
\usepackage{nicefrac}       
\usepackage{microtype}      
\usepackage{lipsum}		
\usepackage{graphicx}
\usepackage{natbib}
\usepackage{doi}
\usepackage{amsmath}
\usepackage{xcolor}

\title{A $q$-Caputo Fractional Generalization of Tsallis Entropy}

\author{ \href{https://orcid.org/0009-0005-4058-6231}{\includegraphics[scale=0.06]{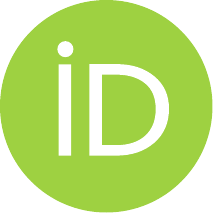}\hspace{1mm}Matias P. Gonzalez}\\
	Departamento de Física\\
	Universidad Católica del Norte\\
	Antofagasta, Chile \\
	\texttt{matias.gonzalez03@alumnos.ucn.cl} \\
	\And
	\href{https://orcid.org/0000-0002-9187-4402}{\includegraphics[scale=0.06]{orcid.pdf}\hspace{1mm}Micolta-Riascos Bayron} \\
	Departamento de Física\\
	Universidad de Antofagasta\\
	Antofagasta, Chile\\
	\texttt{bayron.micolta.riascos@ua.cl} \\
}

\renewcommand{\shorttitle}{\textit{A $q$-Caputo Fractional Generalization of Tsallis Entropy} }

\hypersetup{
  pdftitle={Derivation of a fractional Tsallis entropy using q-Caputo derivative},
  pdfauthor={Matias P. Gonzalez, Bayron Micolta-Riascos},
  pdfsubject={Statistical Mechanics},
  pdfkeywords={Tsallis entropy, q-calculus, fractional calculus, Caputo derivative}}

\begin{document}
\maketitle

\begin{abstract}
In this work we build a $q$-Caputo fractional generalization of the Tsallis entropy functional $S_q$. The construction starts from the already known Jackson derivative representation of $S_q$ and replaces the Jackson derivative with a Caputo type $q$-fractional operator ${}^{C}D_q^{\alpha}$. The parameter $q$ keeps its standard role as the nonextensive index, while the fractional parameter $\alpha$ gives a nonlocal, memory like contribution in the entropy generating procedure. The resulting functional, denoted by $S_q^{\alpha}$, is written as a convergent series of logarithms in the probabilities and is defined for $0<\alpha<1$, $q>0$, and $p_i>0$. We show explicitly that the standard Tsallis entropy is recovered in the local limit $\alpha\to 1$. We then start rewriting $S_q^{\alpha}$ in trace form and analyze its main structural properties. Non-negativity is tested numerically in a binary and ternary probability distributions. These tests show positive values in the respective domains, but they are interpreted just as numerical evidence rather than as a global/formal proof. Concavity is studied through the second derivative of the trace form contribution yielding to a parameter dependent condition. A global statement requires control over the full probability domain. We also discuss Lesche stability and expansibility. Both properties depend on the behavior of the fully resumed functional near its boundary \(p\to 0^+\), because the entropy contains powers of \(\ln p_i\). Finally, we derive the composition rule for statistically independent subsystems. The logarithmic structure produces mixed moments, so the standard Tsallis pseudo-additive law is not directly inherited for \(0<\alpha<1\). As an illustration, we evaluate the proposed entropy in a two-state fractional relaxation process with memory.
\end{abstract}


\section{Introduction}
The Tsallis entropy denoted as $S_q$, has emerged as a useful tool for describing and analyzing systems with long-range interactions, correlations, and fractal phase-space structures \cite{Tsallis:1987eu, Tsallis_2010}. In this work we build a fractional generalization of Tsallis entropy by using $q$-fractional calculus. The approach starts from the Jackson derivative $D_q$ replacing it with a $q$-Caputo type operator, denoted by ${}^{C}D_q^{\alpha}$. This operator provides the analytical basis for the derivation, the study of structural properties and an illustrative application.

The motivation for combining nonextensive statistics with fractional calculus comes from the following: both frameworks describe departures from standard local and additive behavior, but they encode different mechanisms. Nonextensive entropies modify the composition and weighting of probabilities through a deformation parameter, while fractional operators introduce nonlocality through integral kernels and are commonly used in systems with memory, anomalous transport, and non-Markovian relaxation \cite{HilferAnton1995,MetzlerKlafter2000,ScalasGorenfloMainardi2000}. In parallel, several entropy functionals based on fractional calculus have been proposed to incorporate nonlocal or memory like features directly into information measures \cite{Ubriaco2009,Machado2014,LopesMachado2020,FerreiraMachado2019,Bagci2016}. The present construction follows this line of development, but keeps the Tsallis nonextensive parameter $q$ separated from the fractional order $\alpha$. In this way, $q$ controls the nonadditive deformation, whereas $\alpha$ controls the fractional nonlocality of the entropy generating operation.

The paper is organized as follows: In Section~\ref{sec:tsallis_entropy}, we review the standard Tsallis entropy $S_q$, its defining properties, and the procedure by which it can be obtained from the Jackson derivative acting on the generating probabilities $p_i^x$ and evaluated at $x=1$ \cite{Tsallis_2010}. In Section~\ref{sec:q_caputo_derivative}, we introduce the Caputo fractional derivative and its $q$-deformed counterpart, the $q$-Caputo derivative ${}^{C}D_q^{\alpha}$, which provides the main analytical tool of this work. Within the same section, we specify the $q$-Gamma convention and domain in Section~\ref{subsec:q_gamma_convention}, and discuss the physical interpretation of the fractional parameter $\alpha$ in Section~\ref{subsec:physical_interpretation_alpha}. In Section~\ref{sec:derivation_fractional_tsallis}  we apply the $q$-Caputo operator to the Tsallis entropy generating function and derive a fractional generalization of Tsallis entropy, obtaining a closed series representation in terms of the $q$-Gamma function. In Section~\ref{sec:alpha_to_one_limit}, we verify analytically that the standard Tsallis entropy is recovered in the local limit $\alpha\to1$. In Section~\ref{sec:properties}, we examine the main structural properties of the proposed entropy. In particular, Section~\ref{subsec:non_negativity} analyzes its non-negativity domain for two equiprobable microstates and for a ternary non-uniform probability distribution, while Sections~\ref{subsec:concavity}, \ref{subsec:lesche_stability}, and \ref{subsec:pseudo_additivity} are devoted to concavity, Lesche stability, expansibility, and pseudo-additivity, respectively. Finally, in Section~\ref{sec:fractional_relaxation}, we briefly illustrate the use of the proposed entropic functional in the context of fractional relaxation with memory, and Section~\ref{sec:conclusions} summarizes the main results.

\section{Tsallis Entropy}
\label{sec:tsallis_entropy}
Tsallis statistics is based on the generalization of the Boltzmann--Gibbs entropy \(S_{\rm BG}\) into the Tsallis entropy \(S_q\) \cite{Tsallis:1987eu}. This functional is the starting point of the nonextensive formalism and is defined by
\begin{equation}
    S_q \equiv k\,\frac{1 - \sum_i p_i^{\,q}}{q-1},  \label{eq:tsallis-entropy}
\end{equation}
where \(\{p_i\}\) denotes the probabilities of the system microstates, \(k\) is a constant analogous to Boltzmann’s constant, and \(q\) is a real nonextensive parameter. In the limit \(q \to 1\), this expression reduces to the Boltzmann--Gibbs entropy, thereby recovering the standard statistical-mechanical framework.

Tsallis entropy \(S_q\) maintains structural properties of the Boltzmann-Gibbs entropy \(S_{\rm BG}\): for normalized distributions and \(q>0\) it is non-negative, and it attains its maximum for the uniform (equiprobable) distribution. The key difference lies in the composition rule. Whereas \(S_{\rm BG}\) is strictly additive, \(S_q\) is non-additive (pseudo-additive). For two statistically independent subsystems \(A\) and \(B\) with factorized joint probabilities, the total entropy satisfies
\begin{equation}
    S_q(A+B) \;=\; S_q(A) \;+\; S_q(B) \;+\; \frac{1-q}{k}\,S_q(A)\,S_q(B), \label{eq:pseudo_additivity}
\end{equation}
The final term measures the departure from additivity and is governed by the nonextensivity index \(q\); it vanishes as \(q\to1\), recovering the additive Boltzmann--Gibbs limit. This correction does not imply physical correlations between \(A\) and \(B\). Rather, it reflects the effective coupling encoded by the nonextensive entropy, which is useful for modeling systems with long-range interactions, memory effects, or constrained phase-space structures.

A compact route to \eqref{eq:tsallis-entropy} uses the Jackson $q$-derivative, a finite-difference deformation of the ordinary derivative \cite{Tsallis_2010}:
\begin{equation}\label{Jackson derivative}
    D_q f(x) \equiv \frac{f(qx)-f(x)}{qx-x}.
\end{equation}
Identifying the Jackson parameter with the entropic index $q$ and applying $D_q$ to $f_i(x)=p_i^{\,x}$ at $x=1$,
\begin{align}
D_q p_i^{\,x}\Big|_{x=1}
&= \frac{p_i^{\,q}-p_i}{q-1},\\
-\,k\sum_i D_q p_i^{\,x}\Big|_{x=1}
&= -\,k\,\frac{\sum_i p_i^{\,q}-\sum_i p_i}{q-1}
= k\,\frac{1-\sum_i p_i^{\,q}}{q-1}
= S_q,
\end{align}
where we used the normalization $\sum_i p_i=1$. In the limit $q\to 1$ implies $D_1 p_i^{\,x}\big|_{x=1}=p_i\ln p_i$, yielding
the Boltzmann-Gibbs entropy $S_{BG}=-k\sum_i p_i\ln p_i$. The last derivation will be the main road to the fractional generalization of Tsallis entropy.

The Jackson derivative representation of $S_q$ shows that the entropy can be obtained from a deformed derivative acting on the generating family $\sum_i p_i^x$. Therefore, the natural next step is to replace the local $q$-derivative by a fractional $q$-Caputo operation.

\section{$q$-Caputo derivative}
\label{sec:q_caputo_derivative}

The previous section shows that the standard Tsallis entropy can be obtained by applying the Jackson $q$-derivative to the generating family $\sum_i p_i^x$ and evaluating the result at $x=1$. Therefore, in order to construct a fractional extension of this derivation, we first recall the ordinary Caputo derivative and then its $q$-deformed analogue.

Fractional calculus extends integer-order differentiation and integration to non-integer orders. For a positive real order $\alpha>0$, the Riemann--Liouville fractional integral of a function $f(t)$ is defined as
\begin{equation}\label{Fractional integral}
    I^\alpha f(t)
    =
    \frac{1}{\Gamma(\alpha)}
    \int_0^t
    f(\tau)(t-\tau)^{\alpha-1}\,d\tau .
\end{equation}
The Caputo fractional derivative of order $\alpha>0$ is then obtained by first applying an ordinary derivative of integer order $n$ and then a fractional integral of order $n-\alpha$, namely
\begin{equation}\label{Fractional-Caputo-derivative}
    {}^C D^\alpha f(t)
    =
    I^{n-\alpha}D^n f(t)
    =
    \frac{1}{\Gamma(n-\alpha)}
    \int_0^t
    \left[
    \frac{d^n f(\tau)}{d\tau^n}
    \right]
    (t-\tau)^{n-\alpha-1}\,d\tau ,
    \qquad n-1<\alpha<n .
\end{equation}
In particular, for the range used in this work, $0<\alpha<1$, one has $n=1$ and therefore
\begin{equation}\label{Caputo_01}
    {}^C D^\alpha f(t)
    =
    I^{1-\alpha}\frac{d f(t)}{dt}.
\end{equation}
This form makes clear that the Caputo derivative combines a local derivative with a fractional integral kernel. Thus, the resulting operator is nonlocal in the variable on which it acts \cite{zhmakin2022compactintroductionfractionalcalculus}.

The same idea can be implemented in $q$-calculus. The $q$-fractional integral of order $\alpha>0$ is defined as
\begin{equation}\label{q_fractional_integral}
    I_q^\alpha f(t)
    =
    \frac{1}{\Gamma_q(\alpha)}
    \int_0^t
    (t-q\tau)_q^{\alpha-1} f(\tau)\,d_q\tau ,
\end{equation}
where $d_q\tau$ denotes the Jackson $q$-measure and $\Gamma_q$ is the $q$-Gamma function. The corresponding $q$-Caputo derivative of order $\alpha>0$ is
\begin{equation}\label{q_caputo_derivative}
    {}^C D_q^\alpha f(t)
    =
    I_q^{n-\alpha}D_q^n f(t)
    =
    \frac{1}{\Gamma_q(n-\alpha)}
    \int_0^t
    \left[
    D_q^n f(\tau)
    \right]
    (t-q\tau)_q^{n-\alpha-1}\,d_q\tau ,
    \qquad n-1<\alpha<n .
\end{equation}
For the interval $0<\alpha<1$, which is the case considered throughout this manuscript, Eq.~\eqref{q_caputo_derivative} reduces to
\begin{equation}\label{q_caputo_01}
    {}^C D_q^\alpha f(t)
    =
    I_q^{1-\alpha}D_q f(t).
\end{equation}
This expression is the direct $q$-fractional counterpart of Eq.~\eqref{Caputo_01}. It is also the form required to fractionalize the Jackson derivative derivation of the Tsallis entropy.

\subsection{$q$-Gamma convention and domain}
\label{subsec:q_gamma_convention}

In order to avoid ambiguities in the normalization of the $q$-Gamma
function, we specify the convention used throughout this work. For
$0<q<1$ and $\operatorname{Re}(z)>0$, we use the standard $q$-Gamma
function
\begin{equation}\label{eq:q_gamma_product}
    \Gamma_q(z)
    =
    (1-q)^{1-z}
    \frac{(q;q)_\infty}{(q^z;q)_\infty},
\end{equation}
where
\begin{equation}
    (a;q)_\infty
    =
    \prod_{k=0}^{\infty}(1-aq^k)
\end{equation}
is the $q$-Pochhammer symbol \cite{math10010064}. Equivalently, the same convention admits
the integral representation
\begin{equation}\label{eq:q_gamma_integral}
    \Gamma_q(z)
    =
    \int_0^{\frac{1}{1-q}}
    x^{z-1}E_q^{-qx}\,d_qx,
    \qquad 0<q<1,\qquad \operatorname{Re}(z)>0.
\end{equation}
Here
\begin{equation}\label{eq:q_exp_q_number}
    E_q^z
    =
    \sum_{n=0}^{\infty}
    q^{n(n-1)/2}
    \frac{z^n}{[n]_q!},
    \qquad
    [n]_q
    =
    \frac{1-q^n}{1-q}.
\end{equation}

For values $q>1$, we use the standard extension obtained through the
transformation $q\mapsto q^{-1}$,
\begin{equation}\label{eq:q_gamma_extension}
    \Gamma_q(z)
    =
    q^{(z-1)(z-2)/2}\,
    \Gamma_{1/q}(z),
    \qquad q>1.
\end{equation}
With this convention, the recurrence relation
\begin{equation}\label{eq:q_gamma_recurrence}
    \Gamma_q(z+1)
    =
    [z]_q\,\Gamma_q(z),
    \qquad
    [z]_q
    =
    \frac{1-q^z}{1-q},
\end{equation}
holds for $q>0$, $q\neq1$, with the limit $q\to1$ recovering the
ordinary Gamma function. In particular, for integer $m\geq1$,
\begin{equation}\label{eq:q_gamma_integer_identity}
    [m]_q
    =
    \frac{\Gamma_q(m+1)}{\Gamma_q(m)}.
\end{equation}
This is the identity used in the derivation of the fractional entropy.
The case $q=1$ is not included in the definition above and is understood
only as the limiting case in which $\Gamma_q(z)\to\Gamma(z)$ and
$[m]_q\to m$.

For the definitions and properties of the $q$-Gamma function and
$q$-fractional calculus, see Refs.~\cite{math10010064, desole2003integralrepresentationsqgammaqbeta,Annaby2012}.
Here $[n]_q$ is the standard $q$-number and $[n]_q!$ is the corresponding $q$-factorial. The first-order Jackson derivative used in the previous section is recovered as
\begin{equation}\label{Jackson_derivative_section3}
    D_q f(t)
    =
    \frac{f(qt)-f(t)}{qt-t},
\end{equation}
with $d_q f(t)=f(qt)-f(t)$ and $d_qt=(q-1)t$ as the corresponding $q$-differentials. For more details on $q$-fractional calculus and $q$-difference equations, see Refs.~\cite{math10010064, Annaby2012}.

\subsection{Physical interpretation of the fractional parameter $\alpha$}
\label{subsec:physical_interpretation_alpha}

In the present construction, the parameter $\alpha$ should not be interpreted as a second nonextensivity index. The nonextensive deformation is still controlled by $q$, whereas $\alpha$ controls the degree of fractional nonlocality introduced in the entropy generating operation. For $0<\alpha<1$, the relevant operator is
\begin{equation}\label{q_caputo_interpretation}
    {}^C D_q^\alpha f(x)
    =
    I_q^{1-\alpha}D_q f(x).
\end{equation}
Therefore, when this operator acts on the entropy generating function $\sum_i p_i^x$, the entropy evaluated at $x=1$ is not determined only by the local Jackson slope of the generator. Instead, it is determined by a $q$-weighted fractional average of that slope in the auxiliary exponent variable $x$. In this sense, $\alpha$ measures how strongly the entropy functional samples nonlocal information in the entropy generating space.

The limit $\alpha\to1$ suppresses the fractional averaging and recovers the local Jackson derivative construction of the standard Tsallis entropy. $\alpha$ introduces a memory-like weighting in the entropy generating procedure. This interpretation is consistent with the role commonly played by fractional operators in anomalous transport, fractional relaxation, non-Markovian dynamics, and systems with long-time memory \cite{HilferAnton1995,MetzlerKlafter2000,ScalasGorenfloMainardi2000}. From this perspective, the parameters $q$ and $\alpha$ encode different physical effects: $q$ parametrizes nonadditivity or nonextensivity, while $\alpha$ parametrizes the fractional nonlocality of the entropy construction.

Once the mathematical operator and the physical role of $\alpha$ have been specified, we can apply the $q$-Caputo derivative to the same generating family used in the Jackson derivation of the standard Tsallis entropy.

\section{Derivation of the Fractional Tsallis Entropy}
\label{sec:derivation_fractional_tsallis}

We start by rewriting the Tsallis entropy in its Jackson derivative form. 
Throughout this derivation we assume a normalized probability vector with
$p_i>0$, so that $\ln p_i$ is well defined. States with $p_i=0$ can be
treated by a limiting procedure or omitted from the support of the
probability distribution. For $q\neq1$, one has
\begin{equation}
    S_q
    =
    -D_q\sum_{i=1}^{W}p_i^x\Bigg|_{x=1}
    =
    -\sum_{i=1}^{W}D_q p_i^x\Bigg|_{x=1}.
\end{equation}
Using the definition of the Jackson derivative,
\begin{equation}
    D_q p_i^x
    =
    \frac{p_i^{qx}-p_i^x}{qx-x}
    =
    \frac{e^{qx\ln p_i}-e^{x\ln p_i}}{x(q-1)}.
\end{equation}
Expanding both exponentials gives
\begin{equation}
    e^{qx\ln p_i}-e^{x\ln p_i}
    =
    \sum_{m=0}^{\infty}
    \frac{q^m x^m(\ln p_i)^m}{m!}
    -
    \sum_{m=0}^{\infty}
    \frac{x^m(\ln p_i)^m}{m!},
\end{equation}
or, equivalently,
\begin{equation}
    e^{qx\ln p_i}-e^{x\ln p_i}
    =
    \sum_{m=0}^{\infty}
    \frac{(q^m-1)x^m(\ln p_i)^m}{m!}.
\end{equation}
The $m=0$ term vanishes identically because $q^0-1=0$. Therefore, the
series starts effectively at $m=1$, and
\begin{equation}
    D_q p_i^x
    =
    \sum_{m=1}^{\infty}
    \frac{q^m-1}{q-1}
    \frac{(\ln p_i)^m}{m!}
    x^{m-1}.
\end{equation}
Using the standard $q$-number
\begin{equation}\label{m number}
    [m]_q
    =
    \frac{1-q^m}{1-q}
    =
    \frac{q^m-1}{q-1},
    \qquad m\geq1,
\end{equation}
we obtain
\begin{equation}\label{Jackson_series_expansion}
    D_q p_i^x
    =
    \sum_{m=1}^{\infty}
    \frac{[m]_q}{m!}
    (\ln p_i)^m x^{m-1}.
\end{equation}
Hence,
\begin{equation}\label{Tsallis_series_Jackson}
\begin{split}
    S_q
    &=
    -\sum_{i=1}^{W}
    \sum_{m=1}^{\infty}
    \frac{[m]_q}{m!}
    (\ln p_i)^m x^{m-1}
    \Bigg|_{x=1}.
\end{split}
\end{equation}
The interchange between the infinite series and the Jackson derivative is
justified as follows. For $p_i>0$, the function $p_i^x=e^{x\ln p_i}$ is
entire in the auxiliary variable $x$. Therefore, its Taylor expansion
converges uniformly on every compact interval $K\subset(0,\infty)$. Since
the factor $1/[x(q-1)]$ is bounded on such compact intervals for $q\neq1$,
the Jackson differentiated series also converges uniformly in a
neighborhood of the evaluation point $x=1$. For completeness, and in order to justify the series representation obtained above, let us analyze its convergence explicitly. Since the number of accessible microstates is finite, $S_q$ is a finite sum of $W$ infinite series:
\begin{align}
    S_q
    &=
    -\left(
    \sum_{m=1}^{\infty}
    \frac{[m]_q}{m!}
    (\ln p_1)^m x^{m-1}
    +\cdots+
    \sum_{m=1}^{\infty}
    \frac{[m]_q}{m!}
    (\ln p_W)^m x^{m-1}
    \right)\Bigg|_{x=1}.
\end{align}
Therefore, the convergence of $S_q$ follows if the generic infinite series $\displaystyle \sum_{m=1}^{\infty}
    \frac{[m]_q}{m!}
    (\ln p_i)^m x^{m-1}$ is convergent for each fixed $p_i>0$. Denoting its $m$-th term by $a_m$, we apply the ratio test:
\begin{align}
    \lim_{m\to \infty} \left|\frac{a_{m+1}}{a_m}\right|
    &=
    \lim_{m\to \infty}
    \left|
    \frac{[m+1]_q}{(m+1)!}
    (\ln p_i)^{m+1}x^m
    \cdot
    \frac{m!}{[m]_q(\ln p_i)^m x^{m-1}}
    \right|
    \\
    &=
    |x\ln p_i|
    \lim_{m\to\infty}
    \left|
    \frac{1-q^{m+1}}{(m+1)(1-q^m)}
    \right|,
\end{align}
where the relation \eqref{m number} has been used. The last limit vanishes in both regimes. For $0<q<1$, one has $q^m\to0$, so the expression behaves as $1/(m+1)$. For $q>1$, the dominant terms are $q^{m+1}$ and $q^m$, so the expression behaves as $q/(m+1)$. Hence,
\begin{align}
     \lim_{m\to \infty} \left|\frac{a_{m+1}}{a_m}\right|
     =
     0<1.
\end{align}
Thus, the generic series is absolutely convergent for every finite value of $x$. Since $S_q$ is a finite sum of $W$ such convergent series, the standard Tsallis entropy is well defined in this representation. This also provides the necessary justification for using the same term-by-term structure as the starting point for the $q$-Caputo fractional generalization.

We now introduce the $q$-Caputo fractional generalization by replacing the
Jackson derivative with the $q$-Caputo derivative of order $0<\alpha<1$:
\begin{equation}\label{Sqalpha_definition}
    S_q^\alpha
    =
    -\,{}^C D_q^\alpha
    \sum_{i=1}^{W}p_i^x
    \Bigg|_{x=1},
    \qquad 0<\alpha<1.
\end{equation}
For $0<\alpha<1$, the $q$-Caputo derivative is
\begin{equation}
    {}^C D_q^\alpha f(x)
    =
    I_q^{1-\alpha}D_q f(x).
\end{equation}
Therefore,
\begin{equation}
\begin{split}
    S_q^\alpha
    &=
    -I_q^{1-\alpha}D_q
    \sum_{i=1}^{W}p_i^x
    \Bigg|_{x=1}  \\
    &=
    -I_q^{1-\alpha}
    \sum_{i=1}^{W}
    \sum_{m=1}^{\infty}
    \frac{[m]_q}{m!}
    (\ln p_i)^m x^{m-1}
    \Bigg|_{x=1}.
\end{split}
\end{equation}
The term-by-term action of $I_q^{1-\alpha}$ is justified by the
absolute and uniform convergence of the differentiated series in a
compact neighborhood of the evaluation point $x=1$, together with the
linearity of the $q$-fractional integral. Since $p_i^x=e^{x\ln p_i}$ is
analytic for $p_i>0$, the corresponding exponential series provides a
uniformly convergent expansion on compact intervals. Therefore, within
this domain, the $q$-fractional integral can be interchanged with the
infinite summation.

Using the standard identity for the $q$-fractional integral of a power,
\begin{equation}
    I_q^{1-\alpha}x^{m-1}
    =
    \frac{\Gamma_q(m)}
    {\Gamma_q(m+1-\alpha)}
    x^{m-\alpha},
    \qquad m\geq1,
\end{equation}
we obtain the following
\begin{equation}
\begin{split}
    S_q^\alpha
    &=
    -\sum_{i=1}^{W}
    \sum_{m=1}^{\infty}
    \frac{[m]_q}{m!}
    (\ln p_i)^m
    \frac{\Gamma_q(m)}
    {\Gamma_q(m+1-\alpha)}
    x^{m-\alpha}
    \Bigg|_{x=1}.\label{eq33}
\end{split}
\end{equation}
With the next convention of the $q$-Gamma function
\begin{equation}
    \Gamma_q(m+1)
    =
    [m]_q\Gamma_q(m),
    \qquad m\geq1,
\end{equation}
equation \eqref{eq33} becomes
\begin{equation}\label{Our entropy}
    S_q^\alpha
    =
    -\sum_{i=1}^{W}
    \sum_{m=1}^{\infty}
    \frac{\Gamma_q(m+1)}
    {m!\,\Gamma_q(m+1-\alpha)}
    x^{m-\alpha}
    (\ln p_i)^m
    \Bigg|_{x=1}.
\end{equation}
Finally, evaluating at $x=1$, we arrive at the $q$-Caputo fractional
Tsallis entropy
\begin{equation}\label{Our entropylimited}
    S_q^\alpha
    =
    -\sum_{i=1}^{W}
    \sum_{m=1}^{\infty}
    \frac{\Gamma_q(m+1)}
    {m!\,\Gamma_q(m+1-\alpha)}
    (\ln p_i)^m, \qquad 0<\alpha<1,\quad q>0.
\end{equation}
The absence of the $m=0$ term is essential: it follows both from
$[0]_q=0$ in the Jackson expansion and from the fact that the Caputo
derivative annihilates constant contributions. For more details about $q$-calculus and fractional calculus see
\cite{math10010064, Annaby2012}. The expression
\eqref{Our entropylimited} is the central analytical result of this work.
Its consistency must first be tested in the local limit $\alpha\to1$,
where the fractional averaging is removed and the standard Tsallis entropy
should be recovered.

\section{Limit case $\alpha \to 1$}
\label{sec:alpha_to_one_limit}
It is instructive to prove analytically that if we set $\alpha \to 1$ $S_q$ is recovered. Starting from~\eqref{Our entropylimited} we have
\begin{equation}
     S_q^{\alpha\to 1}=-\sum_{i=1}^W\sum_{m=1}^\infty \frac{\Gamma_q(m+1)}{m!\Gamma_q(m)}
    (\ln p_i)^m,
\end{equation}
where the identity
\begin{equation}
    \sum_{m=1}^\infty \frac{[m]_q}{m!}t^m = \frac{e^{qt}-e^t}{q-1},
\end{equation}
is used by defining $t = \ln{p_i}$, the $m = 0$ term in the sum is equal to zero and using $[m]_q = \Gamma_q(m+1)/\Gamma_q(m)$ we get
\begin{equation}
    S_q^{\alpha\to 1} = - \sum_{i = 1}^W \frac{e^{q\ln{p_i}}-e^{\ln{p_i}}}{q-1} = - \sum_{i = 1}^W \frac{p_i^q-p_i}{q-1},
\end{equation}
using $\sum_{i=1}^W p_i = 1$ it simplifies to 
\begin{equation}
    S_q^{\alpha\to 1} = -  \frac{\sum_{i = 1}^W p_i^q-1}{q-1} =   \frac{1 - \sum_{i = 1}^W p_i^q}{q-1},
\end{equation}
which recovers Tsallis entropy~\eqref{eq:tsallis-entropy} assuming natural units $k = 1$. After establishing the recovery of the standard entropy in the limit $\alpha\to1$, we now examine whether the new fractional entropy preserves one of the basic properties of $S_q$, namely non-negativity.

\section{Properties}
\label{sec:properties}

Having obtained the series representation \eqref{Our entropylimited} and verified the recovery of the standard Tsallis entropy in the limit $\alpha\to1$, we now discuss some basic properties of the proposed functional. Since the construction introduces a fractional operation in the entropy generating procedure, the usual properties of $S_q$ are not automatically inherited. In particular, non-negativity, concavity and generalized additivity must be examined explicitly.

In what follows, we work with normalized probability vectors
\begin{equation}
    \mathbf{p}=(p_1,\ldots,p_W),
    \qquad
    p_i>0,
    \qquad
    \sum_{i=1}^{W}p_i=1.
\end{equation}
The restriction $p_i>0$ is imposed because the series representation involves powers of $\ln p_i$. Boundary points with $p_i=0$ are understood as limiting cases. The entropy can be written as a trace-form (see \cite{HanelThurner2011})
\begin{equation}
    S_q^\alpha[\mathbf{p}]
    =
    \sum_{i=1}^{W} h_{q,\alpha}(p_i),\label{eq:trazatypeentropy}
\end{equation}
where
\begin{equation}
    h_{q,\alpha}(p)
    =
    -\sum_{m=1}^{\infty}
    C_m(q,\alpha)\,[\ln p]^m,
    \qquad
    C_m(q,\alpha)
    =
    \frac{\Gamma_q(m+1)}
    {m!\,\Gamma_q(m+1-\alpha)}.\label{eq:relevant}
\end{equation}
This representation is useful because the global properties of $S_q^\alpha$ are controlled by the single probability function $h_{q,\alpha}(p)$ and by the behavior of the coefficients $C_m(q,\alpha)$.

\subsection{Non-negativity analysis}
\label{subsec:non_negativity}
The standard Tsallis entropy is non-negative for normalized probability distributions and $q>0$. In contrast, for the fractional functional $S_q^\alpha$, non-negativity is not guaranteed a priori for every value of $(q,\alpha)$ and every probability vector. The absence of the $m=0$ term in \eqref{Our entropylimited} removes the spurious constant contribution, but the series still contains alternating powers of $\ln p_i<0$. Therefore, the sign of $S_q^\alpha$ must be analyzed from the corrected series beginning at $m=1$.

\begin{figure}[t]
    \centering

    \begin{subfigure}[t]{0.45\textwidth}
        \centering
        \includegraphics[width=\linewidth]{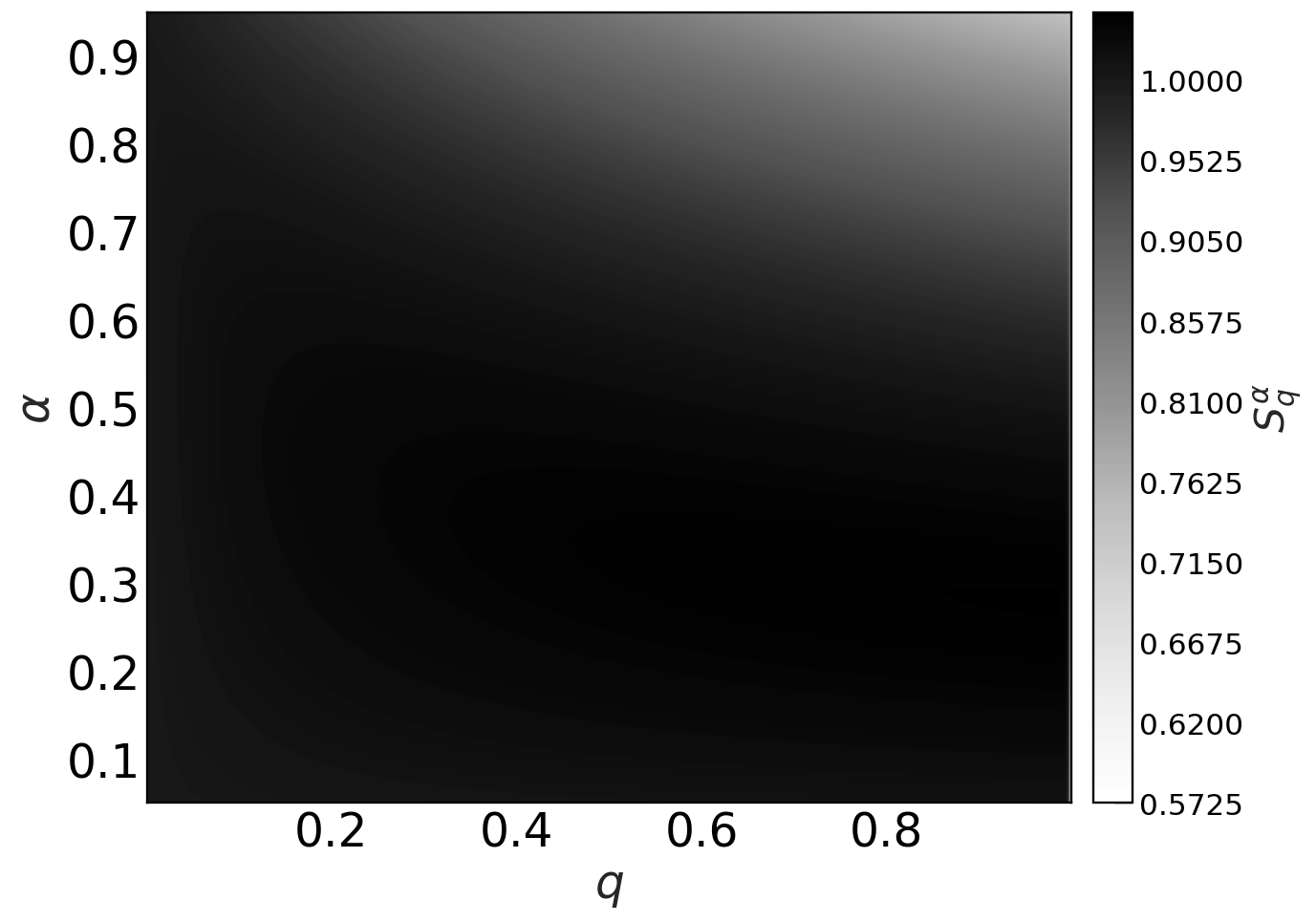}
        \caption{$0<q<1$.}
        \label{fig:Sqalpha_q_less_1}
    \end{subfigure}
    \hfill
    \begin{subfigure}[t]{0.45\textwidth}
        \centering
        \includegraphics[width=\linewidth]{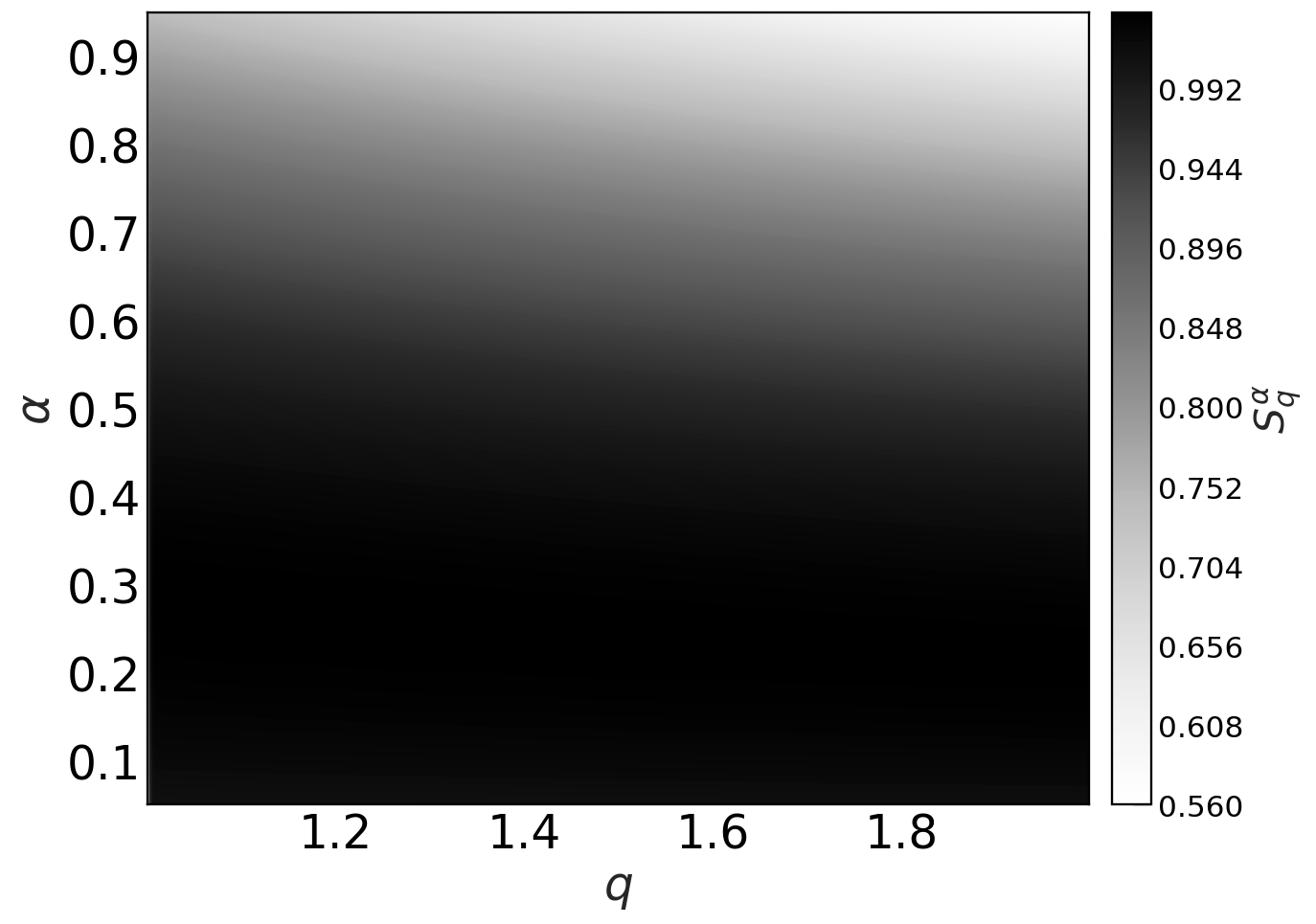}
        \caption{$1<q<2$.}
        \label{fig:Sqalpha_q_greater_1}
    \end{subfigure}

    \caption{\textbf{Binary equiprobable benchmark.}
    Contour maps of $S_q^{\alpha}$ for two equiprobable microstates
    $(W=2,\;p_i=1/2)$, computed from the corrected series representation starting at $m=1$. Panel (a) shows the domain $0<q<1$, while panel (b) shows the extension $1<q<2$ obtained through the corresponding $q$-Gamma continuation. In both panels, the fractional order is varied in the range $0.05\leq\alpha\leq0.95$, and the grayscale bar represents
    the value of $S_q^{\alpha}$.}
    \label{fig:Sqalpha_q_domains}
\end{figure}

We first consider the binary equiprobable benchmark,
\begin{equation}
    W=2,
    \qquad
    p_1=p_2=\frac{1}{2}.
\end{equation}
This case is analytically simple and provides a useful diagnostic for the dependence on $(q,\alpha)$. Figure~\ref{fig:Sqalpha_q_domains} shows that the entropy is non-negative in the plotted parameter domains. This should be interpreted as numerical evidence for the selected ranges, not as a proof of global non-negativity.

\begin{figure}[t]
    \centering
    \includegraphics[width=\textwidth]{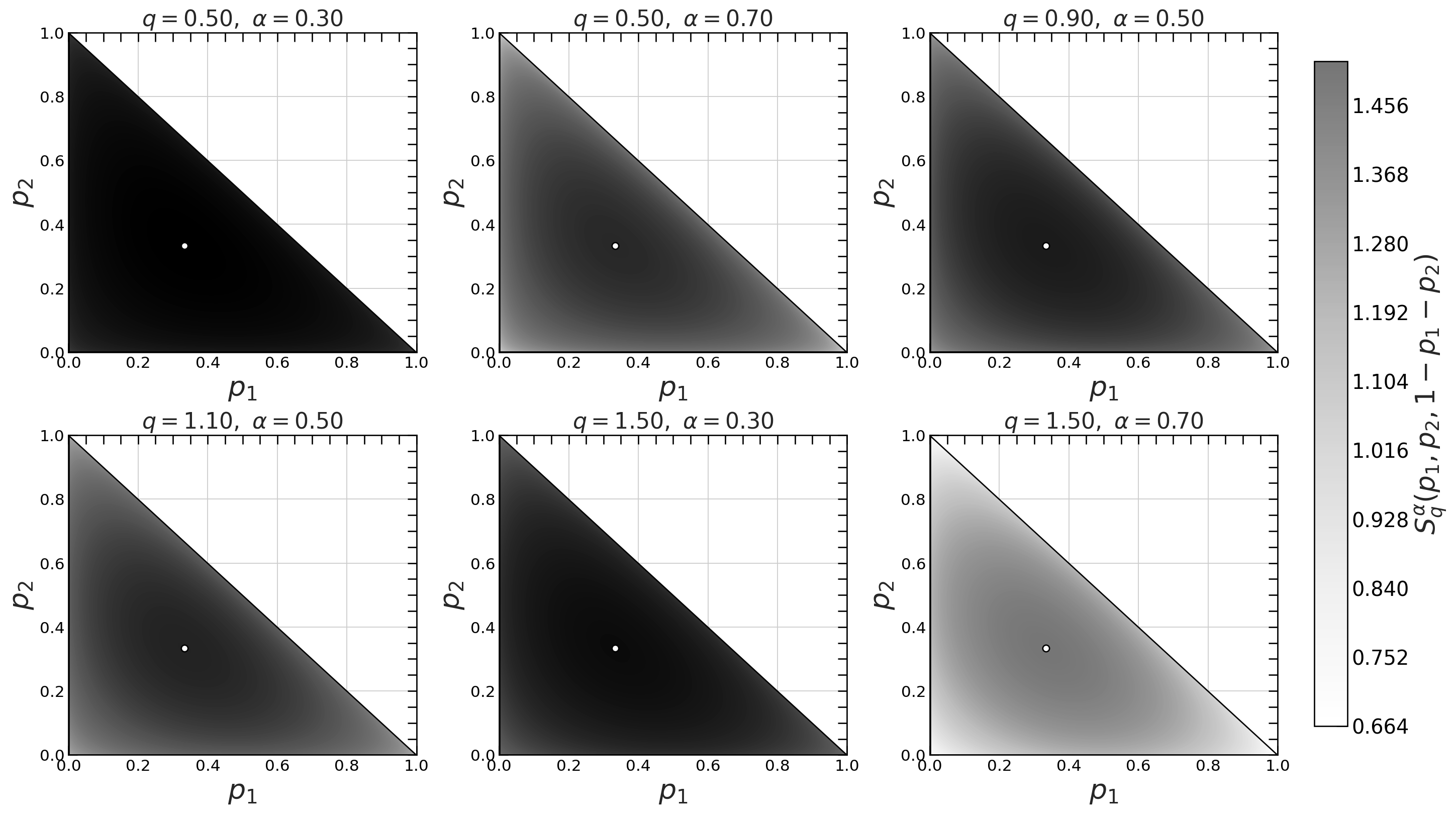}
    \caption{\textbf{Ternary non-uniform probability formulation.}
    Numerical evaluation of $S_q^\alpha(p_1,p_2,p_3)$ over the interior of
    the ternary, with $p_3=1-p_1-p_2$ and $p_i\geq\varepsilon$.
    The panels show representative combinations of $q$ and $\alpha$ in the
    regimes $0<q<1$ and $1<q<2$, including low and high fractional orders
    and values close to the extensive limit $q\to1$.}
    \label{fig:ternary_nonuniform_simplex}
\end{figure}

The binary equiprobable case should be regarded only as a benchmark. To test whether the same sign structure persists away from equiprobability and beyond two states, we also consider a ternary non-uniform probability simplex. We define
\begin{equation}
    \Delta_3
    =
    \left\{
    (p_1,p_2,p_3)\in\mathbb{R}^3:
    p_i>0,\;
    p_1+p_2+p_3=1
    \right\}.
\end{equation}
Numerically, we parametrize the probabilities
\begin{equation}
    p_3=1-p_1-p_2,
    \qquad
    0<p_1<1,
    \qquad
    0<p_2<1,
    \qquad
    p_1+p_2<1.\label{conditions}
\end{equation}
Since the entropy contains powers of $\ln p_i$, the numerical evaluation is performed by imposing a small cutoff
\begin{equation}
p_i\geq\eta.
\end{equation}
The boundary $p_i=0$ is therefore approached only as a limiting case. For the ternary case, the entropy reads
\begin{equation}
    S_q^\alpha(p_1,p_2,p_3)
    =
    -\sum_{i=1}^{3}
    \sum_{m=1}^{\infty}
    \frac{\Gamma_q(m+1)}
    {m!\,\Gamma_q(m+1-\alpha)}
    (\ln p_i)^m.
\end{equation}
The ternary case complements the binary benchmark in two ways. First, it tests non-uniform distributions, where the values of $\ln p_i$ are not all equal. Second, it probes a higher-dimensional probability space. Therefore, it provides a more stringent numerical check of the parameter domains in which $S_q^\alpha$ can be interpreted as a non-negative entropy functional. Figure~\ref{fig:ternary_nonuniform_simplex} shows the numerical evaluation over the $(p_1,p_2)$ simplex for fixed representative values of $q$ and $\alpha$, following the parametrization in Eq.~\eqref{conditions}. The white dot in each panel marks the equiprobable point $p_1=p_2=p_3=1/3$. These plots provide numerical evidence in the sampled domains, but they do not constitute a global analytic proof of non-negativity.

\subsection{Concavity}
\label{subsec:concavity}

An entropic functional $S[\mathbf{p}]$ is concave if, for any two probability distributions $\mathbf{p}$ and $\mathbf{r}$ and for all $\lambda\in[0,1]$, it satisfies
\begin{equation}
    S[\lambda\mathbf{p}+(1-\lambda)\mathbf{r}]
    \geq
    \lambda S[\mathbf{p}]
    +(1-\lambda)S[\mathbf{r}].
\end{equation}
Physically, this means that mixing two statistical states does not decrease the entropy. This property is closely related to thermodynamic stability and to the existence of genuine entropic maxima.

For the standard Tsallis entropy, the trace contribution is
\begin{equation}
    h_q(p)  =
    \frac{p-p^q}{q-1},
\end{equation}
and therefore
\begin{equation}
    h_q''(p) = -q p^{q-2}.
\end{equation}
Thus, for $q>0$ and $0\leq p\leq1$, one has $h_q''(p)\leq0$, showing that the standard Tsallis entropy is concave in this regime \cite{PlastinoPlastino1995}. For the fractional entropy introduced here, the trace-form representation was given in Eq.~\eqref{eq:trazatypeentropy}, with the one-probability function and the coefficients defined in Eq.~\eqref{eq:relevant}. Hence, the concavity of $S_q^\alpha[\mathbf{p}]$ is controlled by the second derivative of $h_{q,\alpha}(p)$.

To compute this derivative, we define
\begin{equation}
    L\equiv\ln p,
    \qquad
    \frac{dL}{dp}=\frac{1}{p}.
\end{equation}
Using Eq.~\eqref{eq:relevant}, the first derivative is
\begin{equation}
\begin{split}
    h_{q,\alpha}'(p)
    &=
    -\sum_{m=1}^{\infty}
    C_m(q,\alpha)
    \frac{d}{dp}\left(L^m\right)  \\
    &=
    -\frac{1}{p}
    \sum_{m=1}^{\infty}
    C_m(q,\alpha)m(\ln p)^{m-1}.
\end{split}
\label{eq:h_first_derivative_concavity}
\end{equation}
For compactness, let us introduce
\begin{equation}
    B_{q,\alpha}(p)
    =
    \sum_{m=1}^{\infty}
    C_m(q,\alpha)m(\ln p)^{m-1}.
\end{equation}
Then Eq.~\eqref{eq:h_first_derivative_concavity} becomes
\begin{equation}
    h_{q,\alpha}'(p)
    =
    -\frac{1}{p}B_{q,\alpha}(p).
\end{equation}
Differentiating once more gives
\begin{equation}
    h_{q,\alpha}''(p)
    =
    \frac{1}{p^2}B_{q,\alpha}(p)
    -
    \frac{1}{p}B_{q,\alpha}'(p).
\end{equation}
The derivative of $B_{q,\alpha}(p)$ is
\begin{equation}
\begin{split}
    B_{q,\alpha}'(p)
    &=
    \sum_{m=1}^{\infty}
    C_m(q,\alpha)m
    \frac{d}{dp}
    \left[(\ln p)^{m-1}\right]  \\
    &=
    \frac{1}{p}
    \sum_{m=2}^{\infty}
    C_m(q,\alpha)m(m-1)(\ln p)^{m-2}.
\end{split}
\end{equation}
The sum starts at $m=2$ because the $m=1$ contribution is proportional to $m-1$ and therefore vanishes. Substituting this result into the expression for $h_{q,\alpha}''(p)$, we obtain
\begin{equation}\label{h''}
\begin{split}
    h_{q,\alpha}''(p)
    &=
    \frac{1}{p^2}
    \sum_{m=1}^{\infty}
    C_m(q,\alpha)m(\ln p)^{m-1}  \\
    &\quad
    -
    \frac{1}{p^2}
    \sum_{m=2}^{\infty}
    C_m(q,\alpha)m(m-1)(\ln p)^{m-2}.
\end{split}
\end{equation}
Separating the $m=1$ term from the first sum and combining the remaining terms, the final expression becomes
\begin{equation}
\begin{split}
    h_{q,\alpha}''(p)
    &=
    \frac{C_1(q,\alpha)}{p^2}
    +
    \frac{1}{p^2}
    \sum_{m=2}^{\infty}
    C_m(q,\alpha)m
    \left[
    (\ln p)^{m-1}
    -
    (m-1)(\ln p)^{m-2}
    \right].
\end{split}
\label{eq:h_second_derivative_concavity}
\end{equation}
Therefore, a sufficient condition for the concavity of the trace-form entropy $S_q^\alpha[\mathbf{p}]$ is
\begin{equation}
    h_{q,\alpha}''(p)\leq0,
    \qquad
    0<p\leq 1.
\end{equation}
Using Eq.~\eqref{eq:h_second_derivative_concavity}, this condition can be written explicitly as
\begin{equation}\label{Concavity condition}
    \frac{C_1(q,\alpha)}{p^2}
    +
    \frac{1}{p^2}
    \sum_{m=2}^{\infty}
    C_m(q,\alpha)m
    \left[
    (\ln p)^{m-1}
    -
    (m-1)(\ln p)^{m-2}
    \right]
    \leq0,
    \qquad
    0<p\leq 1.
\end{equation}
This result shows that concavity is not automatically inherited from the standard Tsallis entropy. Instead, it becomes a parameter-dependent property controlled by the coefficients $C_m(q,\alpha)$ and by the alternating powers of $\ln p$ in the interval $0<p\leq 1$. Thus, concavity is sensitive to the chosen region of parameter space. And is expected to be fully proven in the future.

\subsection{Lesche Stability and Expansibility}
\label{subsec:lesche_stability}
Lesche stability, also referred to as experimental robustness, provides
a criterion for testing if an entropy functional is stable under
small perturbations of the probability distribution
\cite{Lesche1982,KaniadakisScarfone2004,Abe2002Lesche}. Let
\begin{equation}
    \mathbf{p}=(p_1,\ldots,p_W),
    \qquad
    \mathbf{r}=(r_1,\ldots,r_W)
\end{equation}
be two normalized probability distributions in the $W$-state probability simplex. Their distance is measured through the $L^1$ norm
\begin{equation}
    \|\mathbf{p}-\mathbf{r}\|_1
    =
    \sum_{i=1}^{W}|p_i-r_i|.
\end{equation}
An entropy functional $S_W$ is said to be Lesche-stable if, for every entropic tolerance $\epsilon>0$, there exists a probabilistic tolerance $\delta>0$, independent of $W$, such that
\begin{equation}
    \|\mathbf{p}-\mathbf{r}\|_1<\delta
    \quad \Longrightarrow \quad
    \frac{|S_W[\mathbf{p}]-S_W[\mathbf{r}]|}
    {S_W^{\max}}
    <\epsilon ,
    \label{eq:lesche_stability_definition}
\end{equation}
for all $\mathbf{p},\mathbf{r}\in\Delta_W$ and for all $W$. Here
\begin{equation}
    S_W^{\max}
    =
    \max_{\mathbf{p}\in\Delta_W} S_W[\mathbf{p}]
\end{equation}
is the maximum entropy allowed in the $W$-dimensional probability simplex. The normalization by $S_W^{\max}$ is essential because the natural scale of an entropy generally depends on the number of accessible states. Thus, Lesche stability is stronger than ordinary continuity: the same $\delta$ must control the relative entropy variation uniformly for arbitrary $W$.

We now discuss how Lesche stability can be analyzed for our entropic functional. This point requires careful treatment because $S_q^\alpha$ has a nonstandard trace-form structure involving $\ln p_i$ powers. Consequently, the response of the entropy under small perturbations of the probability distribution is controlled not only by the distance $\|\mathbf p-\mathbf r\|_1$, but also by the behavior of the single-probability function near the boundary of the simplex, where $p_i\to0^+$. Therefore, before claiming global Lesche stability, one must distinguish between stability on the interior of the probability simplex and stability on the full simplex including its boundary. The truncated probability simplex is defined as
\begin{equation}
    \Delta_W(\eta)
    =
    \left\{
    \mathbf{p}\in\Delta_W:
    p_i\geq\eta>0,\;
    \sum_{i=1}^{W}p_i=1
    \right\}.
    \label{eq:truncated_simplex}
\end{equation}
This restriction is useful in the present case because the functional
$S_q^{\alpha}$ contains logarithmic powers of the probabilities. Hence,
the boundary points with $p_i=0$ are not part of the natural domain of the
series representation. By imposing a cutoff $p_i\geq\eta$, the analysis is
restricted to the interior of the probability simplex, where all logarithms
are finite.

However, restricted stability on $\Delta_W(\eta)$ is weaker than global
Lesche stability. A global analysis requires control of the full simplex
$\Delta_W$, including the boundary $p_i=0$. Therefore, one must examine the
boundary behavior of the single-probability function
\begin{equation}
    h_{q,\alpha}(p)
    =
    -\sum_{m=1}^{\infty}
    C_m(q,\alpha)(\ln p)^m .
\end{equation}
In particular, the relevant boundary limit is
\begin{equation}
    h_{q,\alpha}(0)
    =
    \lim_{p\to0^+}h_{q,\alpha}(p)
    =
    \lim_{p\to0^+}
    \left[
    -\sum_{m=1}^{\infty}
    C_m(q,\alpha)(\ln p)^m
    \right].
    \label{eq:h_boundary_limit}
\end{equation}
This limit must be evaluated using the complete infinite series. Indeed,
any finite truncation of Eq.~\eqref{eq:trazatypeentropy} generally diverges as $p\to0^+$.
Consequently, a finite boundary value, if it exists, is a property of the
fully resummed functional and cannot be inferred from a finite-order
truncation.

This boundary limit is also directly connected with expansibility. For a
trace-form entropy, expansibility means that adding a zero-probability
state does not change the entropy:
\begin{equation}
    S(p_1,\ldots,p_W,0)
    =
    S(p_1,\ldots,p_W).
\end{equation}
In terms of the single-probability function, this condition is equivalent
to
\begin{equation}
    h_{q,\alpha}(0)=0.
    \label{eq:expansibility_condition}
\end{equation}
Therefore, if the limit in Eq.~\eqref{eq:h_boundary_limit} exists but is
different from zero, the expansibility condition is not satisfied: adding
a null-probability state changes the entropy by a finite amount. If the
limit diverges, the functional cannot be continuously extended to the
boundary of the probability simplex. Thus, the existence and value of
$h_{q,\alpha}(0)$ are necessary ingredients for any global Lesche-stability
analysis.

\subsection{Pseudo-additivity}
\label{subsec:pseudo_additivity}

The pseudo-additive structure of the standard Tsallis entropy is not automatically inherited by the fractional functional $S_q^\alpha$. For two statistically independent subsystems $A$ and $B$, the joint probabilities factorize as
\begin{equation}
    p_{ij}^{A+B}
    =
    p_i^A p_j^B .
\end{equation}
In the standard Tsallis case, this factorization leads directly to the pseudo-additive rule in Eq.~\eqref{eq:pseudo_additivity}. In the present construction, however, the entropy depends on powers of logarithms of the probabilities. Taking the logarithm of the joint probability gives
\begin{equation}
    \ln p_{ij}^{A+B}
    =
    \ln p_i^A+\ln p_j^B .
\end{equation}
Since the functional $S_q^\alpha$ contains all powers $(\ln p_i)^m$ with $m\geq1$, the corresponding term for the composite system becomes
\begin{equation}
    \left(\ln p_{ij}^{A+B}\right)^m
    =
    \left(\ln p_i^A+\ln p_j^B\right)^m .
\end{equation}
Using the binomial expansion, one obtains
\begin{equation}
    \left(
    \ln p_i^A+\ln p_j^B
    \right)^m
    =
    \sum_{\ell=0}^{m}
    \binom{m}{\ell}
    \left(\ln p_i^A\right)^\ell
    \left(\ln p_j^B\right)^{m-\ell}.
\end{equation}
This expression shows the essential difference with respect to the standard Tsallis formalism. In the fractional case, the entropy of the composite system contains mixed logarithmic contributions involving both subsystems. Therefore, the composition rule cannot, in general, be written only in terms of $S_q^\alpha(A)$ and $S_q^\alpha(B)$ through a simple Tsallis-type pseudo-additive law.

To make this structure explicit, we start from the trace-form representation \eqref{eq:trazatypeentropy}. For the composite system one has
\begin{equation}
\begin{split}
    S_q^\alpha(A+B)
    &=
    -\sum_{i=1}^{W_A}
    \sum_{j=1}^{W_B}
    \sum_{m=1}^{\infty}
    C_m(q,\alpha)
    \left(\ln p_{ij}^{A+B}\right)^m
    \\
    &=
    -\sum_{i=1}^{W_A}
    \sum_{j=1}^{W_B}
    \sum_{m=1}^{\infty}
    C_m(q,\alpha)
    \left(
    \ln p_i^A+\ln p_j^B
    \right)^m .
\end{split}
\label{eq:composition_start}
\end{equation}
Substituting the binomial expansion gives
\begin{equation}
\begin{split}
    S_q^\alpha(A+B)
    &=
    -\sum_{i=1}^{W_A}
    \sum_{j=1}^{W_B}
    \sum_{m=1}^{\infty}
    C_m(q,\alpha)
    \sum_{\ell=0}^{m}
    \binom{m}{\ell}
    \left(\ln p_i^A\right)^\ell
    \left(\ln p_j^B\right)^{m-\ell}.
\end{split}
\end{equation}
Since the sums over $i$ and $j$ act on different subsystems, they can be separated:
\begin{equation}
\begin{split}
    S_q^\alpha(A+B)
    &=
    -\sum_{m=1}^{\infty}
    C_m(q,\alpha)
    \sum_{\ell=0}^{m}
    \binom{m}{\ell}
    \left[
    \sum_{i=1}^{W_A}
    \left(\ln p_i^A\right)^\ell
    \right]
    \left[
    \sum_{j=1}^{W_B}
    \left(\ln p_j^B\right)^{m-\ell}
    \right].
\end{split}
\end{equation}
For convenience we define
\begin{equation}
    M_n(A)
    =
    \sum_{i=1}^{W_A}
    \left(\ln p_i^A\right)^n,
    \qquad
    M_n(B)
    =
    \sum_{j=1}^{W_B}
    \left(\ln p_j^B\right)^n,
    \label{eq:log_moments}
\end{equation}
with
\begin{equation}
    M_0(A)=W_A,
    \qquad
    M_0(B)=W_B .
\end{equation}
Using these definitions, the entropy of the composite system can be written as
\begin{equation}
    S_q^\alpha(A+B)
    =
    -\sum_{m=1}^{\infty}
    C_m(q,\alpha)
    \sum_{\ell=0}^{m}
    \binom{m}{\ell}
    M_\ell(A)M_{m-\ell}(B).
    \label{eq:general_log_moment_composition}
\end{equation}

The next step is to separate the endpoint terms $\ell=0$ and $\ell=m$ from the inner sum. Hence
\begin{equation}
\begin{split}
    \sum_{\ell=0}^{m}
    \binom{m}{\ell}
    M_\ell(A)M_{m-\ell}(B)
    &=
    M_0(A)M_m(B)
    +
    M_m(A)M_0(B)
    \\
    &\quad
    +
    \sum_{\ell=1}^{m-1}
    \binom{m}{\ell}
    M_\ell(A)M_{m-\ell}(B).
\end{split}
\end{equation}
Using $M_0(A)=W_A$ and $M_0(B)=W_B$, this becomes
\begin{equation}
\begin{split}
    \sum_{\ell=0}^{m}
    \binom{m}{\ell}
    M_\ell(A)M_{m-\ell}(B)
    &=
    W_A M_m(B)
    +
    W_B M_m(A)
    \\
    &\quad
    +
    \sum_{\ell=1}^{m-1}
    \binom{m}{\ell}
    M_\ell(A)M_{m-\ell}(B).
\end{split}
\end{equation}
Substituting this decomposition into Eq.~\eqref{eq:general_log_moment_composition}, one obtains
\begin{equation}
\begin{split}
    S_q^\alpha(A+B)
    &=
    -W_A
    \sum_{m=1}^{\infty}
    C_m(q,\alpha)M_m(B)
    -
    W_B
    \sum_{m=1}^{\infty}
    C_m(q,\alpha)M_m(A)
    \\
    &\quad
    -
    \sum_{m=1}^{\infty}
    C_m(q,\alpha)
    \sum_{\ell=1}^{m-1}
    \binom{m}{\ell}
    M_\ell(A)M_{m-\ell}(B).
\end{split}
\end{equation}
From the definition of the entropy of each subsystem,
\begin{equation}
    S_q^\alpha(A)
    =
    -\sum_{m=1}^{\infty}
    C_m(q,\alpha)M_m(A),
    \qquad
    S_q^\alpha(B)
    =
    -\sum_{m=1}^{\infty}
    C_m(q,\alpha)M_m(B),
\end{equation}
the first two contributions become
\begin{equation}
    -W_B
    \sum_{m=1}^{\infty}
    C_m(q,\alpha)M_m(A)
    =
    W_B S_q^\alpha(A),
\end{equation}
and
\begin{equation}
    -W_A
    \sum_{m=1}^{\infty}
    C_m(q,\alpha)M_m(B)
    =
    W_A S_q^\alpha(B).
\end{equation}
Finally, the mixed contribution starts at $m=2$, because for $m=1$ the sum $\sum_{\ell=1}^{m-1}$ is empty. Therefore,
\begin{equation}
\begin{split}
    S_q^\alpha(A+B)
    &=
    W_B S_q^\alpha(A)
    +
    W_A S_q^\alpha(B)
    \\
    &\quad
    -
    \sum_{m=2}^{\infty}
    C_m(q,\alpha)
    \sum_{\ell=1}^{m-1}
    \binom{m}{\ell}
    M_\ell(A)M_{m-\ell}(B).
\end{split}
\label{eq:fractional_generalized_composition}
\end{equation}
The last term contains mixed logarithmic moments of the two subsystems. Hence, for $0<\alpha<1$, the fractional entropy satisfies a generalized logarithmic-moment composition rule rather than the standard Tsallis pseudo-additive law. This result is consistent with the physical interpretation of $\alpha$: the fractional order introduces a nonlocal structure in the entropy generating operation, so that the composition rule for $0<\alpha<1$ becomes more involved than the standard Tsallis pseudo-additive law.

\section{Physical example: fractional relaxation with memory}
\label{sec:fractional_relaxation}

To illustrate the possible use of $S_q^\alpha$ in a system with memory effects, let us consider a minimal two-state relaxation process. We take a probability vector
\begin{equation}
    \mathbf{p}(t)=\{p_1(t),p_2(t)\}, 
    \qquad 
    p_1(t)+p_2(t)=1,
\end{equation}
whose relaxation toward equilibrium is governed by a Caputo-type fractional kinetic equation,
\begin{equation}
    {}^{C}D_t^{\beta}\left[p_1(t)-\frac{1}{2}\right]
    =
    -\lambda\left[p_1(t)-\frac{1}{2}\right],
    \qquad 0<\beta\leq1,
\end{equation}
where $\beta$ is the dynamical memory exponent and $\lambda$ is a relaxation rate. Fractional relaxation equations of this type are commonly used to model non-Markovian dynamics and anomalous transport processes \cite{HilferAnton1995,MetzlerKlafter2000}. For the initial condition $p_1(0)=1$, the solution can be written as
\begin{equation}
    p_1(t)=\frac{1}{2}\left[1+E_{\beta}(-\lambda t^\beta)\right],
    \qquad
    p_2(t)=\frac{1}{2}\left[1-E_{\beta}(-\lambda t^\beta)\right].
\end{equation}
The standard Markovian exponential relaxation is recovered for $\beta=1$, whereas $0<\beta<1$ describes a slower relaxation process with memory \cite{MainardiMuraGorenfloStojanovic2007}.

\begin{equation}
    u_\beta(t)\equiv E_{\beta}(-\lambda t^\beta),
\end{equation}
the probabilities are
\begin{equation}
    p_1(t)=\frac{1+u_\beta(t)}{2},
    \qquad
    p_2(t)=\frac{1-u_\beta(t)}{2}.
\end{equation}
Therefore,
\begin{align}
    S_q^\alpha(t)
    &=
    -\sum_{m=1}^{\infty}
    \frac{\Gamma_q(m+1)}
    {m!\,\Gamma_q(m+1-\alpha)}
    \left\{
    \left[
    \ln\left(\frac{1+u_\beta(t)}{2}\right)
    \right]^m
    +
    \left[
    \ln\left(\frac{1-u_\beta(t)}{2}\right)
    \right]^m
    \right\} .
    \label{eq:Sqalpha_fractional_relaxation_evaluated}
\end{align}
Equivalently, the two contributions can be written as
\begin{equation}
    S_q^\alpha(t)=S_{q,+}^{\alpha}(t)+S_{q,-}^{\alpha}(t),
\end{equation}
with
\begin{align}
    S_{q,+}^{\alpha}(t)
    &=
    -\sum_{m=1}^{\infty}
    \frac{\Gamma_q(m+1)}
    {m!\,\Gamma_q(m+1-\alpha)}
    \left[
    \ln\left(\frac{1+u_\beta(t)}{2}\right)
    \right]^m, \\[4pt]
    S_{q,-}^{\alpha}(t)
    &=
    -\sum_{m=1}^{\infty}
    \frac{\Gamma_q(m+1)}
    {m!\,\Gamma_q(m+1-\alpha)}
    \left[
    \ln\left(\frac{1-u_\beta(t)}{2}\right)
    \right]^m .
\end{align}
In the long-time limit, $u_\beta(t)\to0$, and the system approaches the
equiprobable state. Hence,
\begin{equation}
    S_q^\alpha(\infty)
    =
    -2\sum_{m=1}^{\infty}
    \frac{\Gamma_q(m+1)}
    {m!\,\Gamma_q(m+1-\alpha)}
    \left(-\ln 2\right)^m .
\end{equation}
This example separates the two roles of the fractional parameters. The exponent $\beta$ controls the memory of the dynamical relaxation law, while $\alpha$ controls the fractional nonlocality of the entropic functional. Thus, $S_q^\alpha$ gives a possible result of how nonextensive and fractional memory effects modify information content of a relaxing system. This application is also connected with the broader use of generalized entropies in anomalous diffusion and kinetic equations in non linear regimes. In particular, Tsallis statistics has been used to describe anomalous diffusion through nonlinear Fokker-Planck equations \cite{TsallisBukman1996}, while fractional kinetic equations provide a standard framework for memory dominated transport and relaxation \cite{MetzlerKlafter2000}. The present construction combines both ingredients at the level of the entropy functional.

\section{Conclusions}
\label{sec:conclusions}

In this work, we generalized the Tsallis entropy
functional $S_q$ based on a $q$-Caputo fractional scenario. Starting from the
Jackson derivative representation of the standard Tsallis entropy, we replaced
the local $q$-derivative by a Caputo-type $q$-fractional operator. This led to a new
entropy functional, denoted by $S_q^\alpha$, written as a convergent logarithmic
series in the probabilities. In this construction, the nonextensive parameter
$q$ preserves its main usual role, while the fractional order $\alpha$ controls the
nonlocal contribution introduced in the entropy generating operation.

We studied the structural properties of $S_q^\alpha$. The non-negativity of
the functional was explored numerically for representative probability distributions, including the binary equiprobable case and a ternary probability
simplex. These results provide evidence of positive values in the sampled domains, although they do not constitute a global analytic proof. Concavity was analyzed from the trace form representation, leading to a sufficient condition controlled by an alternating logarithmic series. Therefore, concavity is not immediately and automatically inherited from the standard Tsallis entropy, it depends on the
coefficients of the fractional series and on the corresponding parameter space. We also showed that Lesche stability and expansibility are sensitive to the boundary behavior of the functional. A global stability statement requires control of the fully resummed expression in the limit $p_i\to 0^+$, since finite truncations do not generally determine the correct boundary behavior. Similarly, expansibility depends on whether the single probability contribution admits the required limiting value when a zero probability state is added. Thus, these properties must be understood as boundary dependent features of the fractional construction. Finally, we derived the composition rule for statistically independent
subsystems. Since $S_q^\alpha$ depends on powers of logarithms of the
probabilities, the composition law contains mixed logarithmic moments as a main structural feature. As an
illustrative application, we evaluated $S_q^\alpha$ in a two state fractional
relaxation process with memory, separating the role of the dynamical memory
exponent from the fractional order of the entropy functional.

The present formulation should be understood as a first step toward a broader fractional extension of nonextensive statistical mechanics. Future work may strengthen the mathematical control of the boundary behavior, clarify the
thermodynamic interpretation of the fractional order, and explore applications in systems where anomalous relaxation, memory effects, or non-Markovian dynamics are relevant.

\section{Acknowledgements}
MPG acknowledges Vicerrector\'{\i}a de Investigaci\'on y Desarrollo Tecnol\'ogico (VRIDT) at Universidad Cat\'olica del Norte (UCN) for the scientific support provided by N\'ucleo de Investigaci\'on en Simetr\'{\i}as y la Estructura del Universo (NISEU-UCN), Resoluci\'on VRIDT N$^\circ$200/2025.

\section{Funding}
Funding information - not applicable.

\section{Data availability}
Not applicable. This manuscript does not report data generation or analysis.

\section{Competing interests policy
}

No, we declare that the authors have no competing interests as defined by Springer, or other interests that might be perceived to influence the results and/or discussion reported in this paper.

\section{Appendix}
In this appendix we specify the functional setting of the $q$-Caputo operator used in the construction of the fractional entropy. The aim is to state the parameter domain, the Jackson calculus involved, the operator domain, and the precise sense in which the operator acts on the entropy generating function. The classical construction of fractional $q$-integrals and $q$-derivatives can be found in Refs.~\cite{AlSalam1966,Agarwal1969}. Caputo-type $q$-fractional derivatives and related fractional $q$-difference problems are discussed in Refs.~\cite{Annaby2012, StankovicRajkovicMarinkovic2010,AbdeljawadBaleanu2011}.
\subsection{Jackson derivative, Jackson integral, and $q$-fractional integral}

Let $0<q<1$ and let $f:[0,a]\to\mathbb{R}$ be a function defined on a finite interval. For $x\neq0$, the Jackson derivative is defined by
\begin{equation}
    D_q f(x)
    =
    \frac{f(qx)-f(x)}{qx-x}
    =
    \frac{f(x)-f(qx)}{(1-q)x},
    \qquad 0<x\leq a,
    \label{app:jackson_derivative}
\end{equation}
provided that the right-hand side exists. At $x=0$, $D_q f(0)$ is understood through the corresponding limit whenever that limit exists. The Jackson integral over $[0,x]$ is
\begin{equation}
    \int_0^x g(t)\,d_qt
    =
    (1-q)x
    \sum_{k=0}^{\infty}
    q^k g(xq^k),
    \qquad 0<x\leq a,
    \label{app:jackson_integral}
\end{equation}
whenever the series on the right-hand side converges. In particular, if $g$ is bounded on $[0,x]$, then
\begin{equation}
    \left|
    \int_0^x g(t)\,d_qt
    \right|
    \leq
    (1-q)x\sum_{k=0}^{\infty}q^k\|g\|_\infty
    =
    x\|g\|_\infty,
\end{equation}
and the Jackson integral is finite.

For $\mu>0$, the left-sided $q$-fractional integral of order $\mu$ is defined by
\begin{equation}
    I_{q,0+}^{\mu}g(x)
    =
    \frac{1}{\Gamma_q(\mu)}
    \int_0^x
    (x-qt)_q^{\mu-1}g(t)\,d_qt,
    \qquad 0<x\leq a,
    \label{app:q_fractional_integral}
\end{equation}
provided that the Jackson integral on the right-hand side is convergent. Here $(x-qt)_q^{\mu-1}$ denotes the standard $q$-shifted power used in fractional $q$-calculus \cite{math10010064, Annaby2012}.

The $q$-Gamma function is taken with the convention
\begin{equation}
    \Gamma_q(z)
    =
    (1-q)^{1-z}
    \frac{(q;q)_\infty}{(q^z;q)_\infty},
    \qquad
    0<q<1,
    \qquad
    \operatorname{Re}(z)>0,
    \label{app:q_gamma_product}
\end{equation}
where $(a;q)_\infty=\prod_{k=0}^{\infty}(1-aq^k)$ is the $q$-Pochhammer symbol. With this normalization,
\begin{equation}
    \Gamma_q(z+1)=[z]_q\Gamma_q(z),
    \qquad
    [z]_q=\frac{1-q^z}{1-q}.
    \label{app:q_gamma_recurrence}
\end{equation}
Integral representations and normalization conventions for the $q$-Gamma function are discussed in Ref.~\cite{desole2003integralrepresentationsqgammaqbeta}.

\subsection{Left-sided $q$-Caputo derivative}

For $0<\alpha<1$, the left-sided $q$-Caputo derivative is defined by applying first the Jackson derivative and then the left-sided $q$-fractional integral of order $1-\alpha$:
\begin{equation}
    {}^{C}D_{q,0+}^{\alpha}f(x)
    =
    I_{q,0+}^{1-\alpha}D_qf(x).
    \label{app:q_caputo_definition}
\end{equation}
Equivalently,
\begin{equation}
    {}^{C}D_{q,0+}^{\alpha}f(x)
    =
    \frac{1}{\Gamma_q(1-\alpha)}
    \int_0^x
    (x-qt)_q^{-\alpha}
    D_qf(t)\,d_qt.
    \label{app:q_caputo_integral_form}
\end{equation}
Thus, a natural domain for the operator is
\begin{equation}
    \mathcal{D}_{q,\alpha}
    =
    \left\{
    f:[0,a]\to\mathbb{R}:
    D_q f \ \text{exists and } I_{q,0+}^{1-\alpha}D_q f(x)
    \ \text{converges for } x\in(0,a]
    \right\}.
    \label{app:q_caputo_domain}
\end{equation}
The subscript $0+$ indicates that the operator is left-sided and that the lower terminal is fixed at $0$. In the main text we use the shorter notation ${}^{C}D_q^{\alpha}$ for ${}^{C}D_{q,0+}^{\alpha}$. Since $D_qC=0$ for a constant $C$, one has ${}^{C}D_{q,0+}^{\alpha}C=0$. This Caputo-type property explains why the constant term in the exponential expansion does not contribute to the entropy series.

\subsection{Action on the entropy generating function}

In this work, the operator acts on the entropy generating function
\begin{equation}
    F_{\mathbf{p}}(x)
    =
    \sum_{i=1}^{W}p_i^x,
    \qquad
    p_i>0,
    \qquad
    \sum_{i=1}^{W}p_i=1.
    \label{app:entropy_generator}
\end{equation}
The variable $x$ is an auxiliary exponent variable, not a physical time or spatial coordinate. Since $p_i^x=e^{x\ln p_i}$, each term is analytic in $x$ for $p_i>0$. Therefore, $F_{\mathbf p}(x)$ is analytic and its Taylor expansion converges uniformly on compact intervals $K\subset(0,\infty)$, in particular in a neighborhood of the evaluation point $x=1$.

Using the Jackson derivative,
\begin{equation}
    D_qp_i^x
    =
    \frac{p_i^{qx}-p_i^x}{qx-x},
\end{equation}
and expanding the exponentials gives
\begin{equation}
    p_i^{qx}-p_i^x
    =
    \sum_{m=0}^{\infty}
    \frac{(q^m-1)x^m(\ln p_i)^m}{m!}.
\end{equation}
The term $m=0$ vanishes because $q^0-1=0$. Hence,
\begin{equation}
    D_qp_i^x
    =
    \sum_{m=1}^{\infty}
    \frac{[m]_q}{m!}
    (\ln p_i)^m x^{m-1}.
    \label{app:jackson_series}
\end{equation}
The uniform convergence of the exponential series on compact intervals, together with the boundedness of $1/[x(q-1)]$ around $x=1$ for $q\neq1$, justifies the term-by-term Jackson differentiation in the domain used here.

Applying ${}^{C}D_{q,0+}^{\alpha}=I_{q,0+}^{1-\alpha}D_q$ term by term and using
\begin{equation}
    I_{q,0+}^{1-\alpha}x^{m-1}
    =
    \frac{\Gamma_q(m)}
    {\Gamma_q(m+1-\alpha)}
    x^{m-\alpha},
    \qquad m\geq1,
    \label{app:q_fractional_power}
\end{equation}
we obtain, with $\Gamma_q(m+1)=[m]_q\Gamma_q(m)$,
\begin{equation}
    {}^{C}D_{q,0+}^{\alpha}p_i^x
    =
    \sum_{m=1}^{\infty}
    \frac{\Gamma_q(m+1)}
    {m!\,\Gamma_q(m+1-\alpha)}
    (\ln p_i)^m x^{m-\alpha}.
\end{equation}
Consequently,
\begin{equation}
    S_q^\alpha[\mathbf{p}]
    =
    -{}^{C}D_{q,0+}^{\alpha}F_{\mathbf{p}}(x)\Big|_{x=1}
    =
    -\sum_{i=1}^{W}
    \sum_{m=1}^{\infty}
    \frac{\Gamma_q(m+1)}
    {m!\,\Gamma_q(m+1-\alpha)}
    (\ln p_i)^m.
    \label{app:fractional_entropy_final}
\end{equation}
This is the coefficient representation used in the main text.

\subsection{Remark on the $q>1$ regime}

The operatorial definition above is stated in the standard Jackson integral setting $0<q<1$. The sector $q>1$ considered in the numerical analysis is introduced at the level of the final series coefficients through the standard continuation
\begin{equation}
    \Gamma_q(z)
    =
    q^{(z-1)(z-2)/2}
    \Gamma_{1/q}(z),
    \qquad q>1.
    \label{app:q_gamma_gt1}
\end{equation}
Accordingly, the results for $q>1$ should be interpreted as the analytic continuation of the coefficient representation in Eq.~\eqref{app:fractional_entropy_final}, not as an independent operatorial construction based directly on the Jackson integral for $q>1$.
\bibliographystyle{unsrt}
\bibliography{references}  






\end{document}